\documentclass[11pt]{article}
\usepackage[a4paper,margin=1in]{geometry}
\usepackage{amsmath,amssymb,amsthm}
\usepackage{graphicx}
\usepackage{booktabs}
\usepackage{array}
\usepackage[round]{natbib}
\usepackage{algorithm}
\usepackage{algpseudocode}
\usepackage{microtype}
\usepackage{xcolor}
\usepackage[colorlinks=true,linkcolor=blue!50!black,citecolor=blue!50!black,urlcolor=blue!50!black]{hyperref}

\theoremstyle{plain}
\newtheorem{theorem}{Theorem}
\newtheorem{lemma}{Lemma}
\newtheorem{proposition}{Proposition}
\newtheorem{corollary}{Corollary}
\newtheorem{assumption}{Assumption}
\theoremstyle{definition}
\newtheorem{definition}{Definition}
\theoremstyle{remark}
\newtheorem{remark}{Remark}

\title{Koopman Greedy Spiking Control for Nonlinear Systems\\ under Firing-Rate Limits}
\author{Nader Meskin\thanks{Department of Electrical Engineering, Qatar University, Doha, Qatar. E-mail: \texttt{nmeskin@gmail.com}.}}
\date{Preprint submitted to \emph{Automatica}, October 2026}

\begin{document}
\maketitle

\begin{abstract}
Neuromorphic controllers interact with physical systems through fixed-amplitude spikes whose timing conveys control information. Existing Lyapunov-based greedy spiking controllers guarantee stability but are restricted to linear systems and neglect firing-rate limits. We extend greedy spiking control to nonlinear systems using Koopman operator theory. For observable dictionaries invariant under differentiation along the input direction, spike-induced state translations admit an exact linear representation in lifted coordinates, enabling exact modeling of spiking jump dynamics. This allows firing decisions to be formulated as comparisons of quadratic Lyapunov functions in the lifted space. We prove practical stability of a clocked, rate-limited controller with an explicit ultimate bound and develop a data-driven design based on Koopman eigenfunctions learned from unforced data by generator-based regression. The weights of the Lyapunov function are synthesized by a linear program that enforces a sampled relaxation of the rate-limited control-Lyapunov condition. Applied to an inverted pendulum, the learned eigenfunctions recover the system energy, and the proposed controller achieves swing-up and stabilization from nearly all initial conditions despite severe actuation limits.
\end{abstract}


\section{Introduction}

Neuromorphic engineering develops sensors, processors, and actuators that communicate through asynchronous fixed-amplitude spikes inspired by biological nervous systems~\citep{Mead1990}. Neuromorphic processors such as Loihi consume energy primarily when spikes occur and offer extremely low-latency computation~\citep{Davies2018}. These advantages have motivated \emph{neuromorphic control}, where feedback is closed through spike timing rather than continuously valued signals~\citep{Sepulchre2022}.

Recent years have seen rapid progress in the control-theoretic foundations of neuromorphic control. Building on event-triggered control~\citep{AstromBernhardsson2002,HeemelsJohanssonTabuada2012} and hybrid systems theory~\citep{GoebelSanfeliceTeel2012}, several architectures with stability guarantees have emerged. These include hybrid-systems analyses of simple neuromorphic controllers~\citep{Petri2024CDC}, emulation-based approaches that approximate analog controllers with spiking networks~\citep{Petri2025arXivEmu,Eilers2026ECC,Petri2026arXivComm}, Lyapunov-based greedy controllers that fire only when a spike decreases a Lyapunov function~\citep{Heemels2026EJC,Klip2026IFAC}, and spike-coding or oscillatory control architectures inspired by computational neuroscience~\citep{Boerlin2013,Agliati2025,Petri2025Rhythmic}. Despite these advances, most certified designs are restricted to linear plants and quadratic Lyapunov functions. Moreover, they typically neglect firing-rate limitations imposed by neuromorphic hardware through clocked implementations or refractory periods. Under such constraints, the achievable average actuation may be insufficient to counter destabilizing nonlinear dynamics away from equilibrium, making genuinely nonlinear designs necessary.

The Koopman operator provides a systematic framework for nonlinear analysis and control by representing nonlinear dynamics through a linear operator acting on observable functions of the state~\citep{Koopman1931,Mezic2005,Budisic2012,Brunton2022}. Finite-dimensional approximations can be learned from data using extended dynamic mode decomposition and its generator variants~\citep{Williams2015,Klus2020}, and have been used for predictive and robust control~\citep{Korda2018,Strasser2023}, while Koopman eigenfunctions have been used for stability analysis and control design~\citep{Mauroy2016,Kaiser2021}. For control-affine systems, however, Koopman models are generally bilinear and exact only under restrictive invariant-subspace conditions involving both drift and input vector fields~\citep{Goswami2017}. Although recent work has explored control Lyapunov functions for bilinear Koopman models~\citep{Hanna2026}, the Koopman framework has not yet been used in the context of neuromorphic spiking control.

This paper exploits a key observation: spiking control is particularly well suited to Koopman lifting. A spike acts as an impulse that translates the state along the input direction rather than through the continuous vector field. If the observable dictionary is invariant under differentiation along this direction, a considerably weaker requirement than the standard invariant-subspace condition, the translation induced by every spike is represented exactly by a linear operator in the lifted coordinates. Polynomial and trigonometric dictionaries do satisfy this property. Consequently, the jump dynamics of a spiking closed-loop system can be modeled without approximation, and only the flow between spikes requires approximation or bounding.

Despite the complementary strengths of neuromorphic control and Koopman-based modeling, their integration remains largely unexplored. Existing neuromorphic controllers with formal guarantees are predominantly developed for linear systems and do not explicitly address nonlinear dynamics under firing-rate constraints. Conversely, Koopman-based control methods provide powerful tools for nonlinear analysis and synthesis but have been formulated primarily for continuously actuated systems rather than event-driven spiking architectures. As a result, there is currently no unified framework that combines the computational efficiency of sparse spike-based actuation with Koopman-based nonlinear representations and rigorous closed-loop stability guarantees. This paper addresses this gap by exploiting the exact Koopman representation of spike-induced state translations and using it to develop a data-driven greedy spiking control framework for nonlinear systems subject to firing-rate limitations. The main contributions are as follows.

\begin{enumerate}
\item \emph{Exact Koopman representation of spikes.}
We show that spike-induced state translations admit an exact linear representation in lifted coordinates whenever the observable dictionary is invariant under differentiation along the input direction. As a result, firing decisions of greedy spiking controllers with quadratic lifted Lyapunov functions reduce to exact comparisons of quadratic forms.

\item \emph{Koopman Greedy Spiking Control (KGSC) under firing-rate limits.}
For a clocked controller that permits at most one spike per sampling interval, we establish practical stability on certified Lyapunov sublevel sets and derive an explicit ultimate bound. The analysis relies on a bounded-input control Lyapunov condition determined by the maximal achievable average actuation.

\item \emph{Data-driven eigenfunction-based design.}
We construct Lyapunov functions using only Koopman eigenfunctions associated with neutral and unstable modes, which are identified from unforced data through generator extended dynamic mode decomposition.

\item \emph{Convex Lyapunov synthesis.}
The Lyapunov weights are obtained from a linear program enforcing a sampled relaxation of the rate-limited control Lyapunov condition. This convex formulation plays a role analogous to Riccati-based synthesis in the linear greedy framework.

\end{enumerate}

The paper is organised as follows. Section~\ref{sec:setting} states the problem and recalls the
necessary background. Section~\ref{sec:lifting} derives the lifted Koopman model of the spiking system,
whose jump map is exact, and defines the greedy firing rule on the lifted state. Section~\ref{sec:stability}
gives the stability analysis under firing-rate limits, and Section~\ref{sec:design} the
eigenfunction-based design and the convex synthesis of the Lyapunov function. Section~\ref{sec:sim} reports the pendulum case study, and Section~\ref{sec:conclusion} concludes and outlines
future work.

\section{Problem setting and preliminaries}\label{sec:setting}
\subsection{Notation}
 For a function
$h:\mathcal{X}\to\mathbb{R}$ and a vector field $g$, $\mathcal L_gh(x):=\nabla h(x)^\top g(x)$ is the Lie
derivative, and $\mathcal{L}_B h(x):=\nabla h(x)^\top B$ for a constant vector $B$. $\Phi_\vartheta(x)$ is the
solution at time $\vartheta\ge0$ of $\dot x=f(x)$ from the initial state $x$ (the flow of the drift). $x(t^-)$ and
$x(t^+)$ are the left and right limits of $x$ at $t$. $I_m$ is the
$m\times m$ identity matrix, and $\mathrm{sgn}$ is the sign function. $P\succ0$ means that $P$ is
symmetric positive definite.

\subsection{Spiking control problem}

Consider the nonlinear control-affine system \begin{equation}\label{eq:plant} \dot x(t)=f(x(t))+B\,u(t), \end{equation} where $x(t)\in\mathcal{X}\subseteq\mathbb{R}^n$ is the state, $u(t)\in\mathbb{R}$ is the control input, $f:\mathcal{X}\to\mathbb{R}^n$ is locally Lipschitz with $f(0)=0$, and $B\in\mathbb{R}^n$ is a constant input direction. In the neuromorphic-control setting, the input is generated as \begin{equation}\label{eq:input} u(t)=\sum_{i\ge1}s_i\,\alpha\,\delta(t-t_i), \end{equation} where $\alpha>0$ is the fixed spike amplitude, $t_i$ is the time of the $i$th spike, $s_i\in\{-1,1\}$ denotes the spike sign, and $\delta$ is the Dirac impulse. Between spikes, the state evolves according to the drift dynamics $\dot x(t)=f(x(t))$, whereas each spike induces the jump \begin{equation}\label{eq:jump} x(t_i^+)=x(t_i^-)+s_i\,\alpha B . \end{equation} Trajectories are taken to be left-continuous, $x(t)=x(t^-)$, so that $x(t_i)$ denotes the state immediately before the $i$th spike.

\par As is typical for digital neuromorphic hardware, spikes may only be generated at the ticks of a clock. Let $\tau>0$ denote the clock period and define $t_k:=k\tau$, $k=0,1,2,\ldots$. At each tick, the controller selects  $s(t_k)\in\{-1,0,1\}.$  If $s(t_k)\neq0$, a spike of sign $s(t_k)$ is applied at $t_k$; otherwise, no spike is generated. Hence every spike time coincides with a clock tick, at most one spike can be fired per tick, and the maximum firing rate is $1/\tau$. Consequently, over any interval of length $T$, at most $T/\tau+1$ spikes may occur, implying \[ \int_t^{t+T}|u(r)|\,\mathrm dr \le \alpha\left(\frac{T}{\tau}+1\right). \] Therefore, the long-term average control magnitude is bounded by \begin{equation}\label{eq:ubar} \bar u:=\frac{\alpha}{\tau}. \end{equation}

\par Under the left-continuous convention, $x(t_k)$ denotes the state just before the control decision at time $t_k$, while the post-spike state is \[ x(t_k^+)=x(t_k)+s(t_k)\alpha B. \] Since no spikes occur on $(t_k,t_{k+1})$, the state at the next clock tick satisfies $x(t_{k+1}) = \Phi_\tau\!\left(x(t_k^+)\right)$.

\emph{Problem.} Given $f$, $B$, $\alpha$, and $\tau$, construct a state-feedback law that determines the spike decision $s(t_k)\in\{-1,0,1\}$ at each clock tick from the current state $x(t_k)$, such that the closed-loop system admits a certified practically stable operating region. The design should provide explicit bounds on the region of attraction and the ultimate invariant set arising from the firing-rate limitation, while requiring only computationally inexpensive operations compatible with real-time neuromorphic implementation.

\subsection{Koopman operator, lifting and generator EDMD} Consider first the unforced system $ \dot x(t)=f(x(t))$. The Koopman semigroup associated with the drift acts on observables, i.e.\ functions $h:\mathcal X\to\mathbb R$, through composition with the flow map, $(\mathcal K^\vartheta h)(x) := h(\Phi_\vartheta(x))$, $\vartheta\ge0$. Although the state dynamics are generally nonlinear, the Koopman operator is linear in the observable $h$~\citep{Mezic2005,Budisic2012} and its infinitesimal generator is, for continuously differentiable $h$,
\begin{equation}\label{eq:gen}
\mathcal{L}_f h(x) := \lim_{\vartheta\to0^+} \frac{h(\Phi_\vartheta(x))-h(x)}{\vartheta} = \nabla h(x)^\top f(x),
\end{equation} so that $\frac{\mathrm d}{\mathrm dt}h(x(t)) = \mathcal{L}_f h(x(t))$  along unforced trajectories. An eigenfunction $\varphi$ associated with eigenvalue $\lambda$ satisfies  $\mathcal{L}_f\varphi(x)=\lambda\varphi(x)$, and therefore evolves according to $\varphi(x(t)) = e^{\lambda t}\varphi(x(0))$. Hence, Koopman eigenfunctions provide nonlinear coordinates whose dynamics are exactly linear.

To obtain a finite-dimensional linear representation, a dictionary of $N$ user-chosen observables is selected, \begin{equation}\label{eq:psi} \psi(x) = \big(\psi_1(x),\dots,\psi_N(x)\big)^\top, \qquad \psi_i:\mathcal X\to\mathbb R, \end{equation} where the $\psi_i$ are continuously differentiable and linearly independent, for example monomials or trigonometric functions~\citep{Williams2015}. Defining the lifted state $ y(t):=\psi(x(t))\in\mathbb R^N$,  the unforced dynamics satisfy  $\dot y(t) = \mathcal{L}_f\psi(x(t))$. If the dictionary  $\mathrm{span}\{\psi_1,\ldots,\psi_N\}$ is invariant under $\mathcal{L}_f$, then there exists a matrix $A\in\mathbb R^{N\times N}$ such that \[ \mathcal{L}_f\psi(x)=A\psi(x) \] for all $x$, and the lifted state obeys the exact linear model \[ \dot y(t)=Ay(t). \] In general, this invariance condition does not hold, and $A$ is chosen so that $A\psi$ approximates $\mathcal{L}_f\psi$. Defining the residual \begin{equation}\label{eq:res} r(x):=\mathcal{L}_f\psi(x)-A\psi(x), \end{equation} the lifting approximation error is quantified by \[ \dot y(t)=Ay(t)+r(x(t)). \]

\par \emph{Generator EDMD.} Given samples $x^{(j)}$, $j=1,\ldots,M$, together with estimates of \[ \mathcal{L}_f\psi(x^{(j)}) = \nabla\psi(x^{(j)})f(x^{(j)}), \] obtained either from a known model or from finite-difference approximations along measured trajectories, generator extended dynamic mode decomposition (gEDMD)~\citep{Klus2020} determines $A$ by solving the least-squares problem \begin{equation}\label{eq:gedmd} A := \arg\min_{A} \sum_{j=1}^{M} \big\| \mathcal{L}_f\psi(x^{(j)}) - A\psi(x^{(j)}) \big\|^2 = \dot\Psi\,\Psi^\dagger, \end{equation} where the columns of $\Psi$ and $\dot\Psi$ are $\psi(x^{(j)})$ and $\mathcal{L}_f\psi(x^{(j)})$, respectively, and $\dagger$ denotes the Moore--Penrose pseudoinverse.

For the continuously actuated system \[ \dot x(t)=f(x(t))+Bv(t), \] the lifted dynamics satisfy \[ \dot y(t) = \mathcal{L}_f\psi(x(t)) + \mathcal{L}_B\psi(x(t))\,v(t), \] where $ \mathcal{L}_B\psi(x):=\nabla\psi(x)B$.  If the dictionary span is invariant under both $\mathcal{L}_f$ and $\mathcal{L}_B$, the lifted dynamics admit an exact bilinear representation \[ \dot y=Ay+Ny\,v \] for suitable matrices $A$ and $N$. This corresponds to the invariant subspace condition of \citet{Goswami2017}, which is generally difficult to satisfy for nonlinear control-affine systems. Consequently, both the drift and input terms are typically approximated, introducing modeling errors in the lifted dynamics.

\par The preceding discussion concerns continuously actuated systems. For such systems, exact finite-dimensional Koopman representations usually require invariance with respect to both the drift and input operators, a restrictive condition that is rarely satisfied. In contrast, the spike-based inputs considered in this paper act through instantaneous state translations rather than continuous forcing. As shown in the next section, this structure permits an exact Koopman representation of spike-induced jumps under substantially weaker conditions on the observable dictionary.
\section{Lifted Koopman model of the spiking system and greedy firing rule}\label{sec:lifting}
For the spiking system~\eqref{eq:plant}, the input affects the state only through the jumps~\eqref{eq:jump}, which translate the state along the direction $B$. Such translations are generated by the operator $\mathcal{L}_B$. We therefore say that the dictionary span $\mathrm{span}\{\psi_1,\ldots,\psi_N\}$ is \emph{$\mathcal{L}_B$-invariant} if there exists a matrix $L\in\mathbb R^{N\times N}$ such that \begin{equation}\label{eq:LBinv}
\mathcal{L}_B\psi(x)=L\,\psi(x) \qquad\text{for all }x\in\mathcal X.
\end{equation}
This condition corresponds to invariance under the input operator in the bilinear Koopman framework of \citet{Goswami2017}, but does not require invariance under the drift operator $\mathcal{L}_f$.

\begin{lemma}[Exact jump lifting]\label{lem:jump}
Suppose that the dictionary span is $\mathcal{L}_B$-invariant with matrix $L$. Then \begin{equation}\label{eq:exactjump}
\psi(x+\sigma B) = e^{\sigma L}\psi(x) \qquad \text{for all }x\in\mathcal X \text{ and }\sigma\in\mathbb R.
\end{equation}
Consequently, the spike-induced jump~\eqref{eq:jump} is represented exactly in the lifted coordinates as \[ \psi(x(t_i^+)) = e^{s_i\alpha L} \psi(x(t_i^-)). \] \end{lemma}
\begin{proof}
Fix $x\in\mathcal X$ and define $\chi(\sigma):=\psi(x+\sigma B)$. By the chain rule and~\eqref{eq:LBinv}, \[ \frac{\mathrm d}{\mathrm d\sigma}\chi(\sigma) = \mathcal{L}_B\psi(x+\sigma B) = L\,\chi(\sigma), \] with initial condition $\chi(0)=\psi(x)$. Hence $\chi$ satisfies a linear time-invariant differential equation whose unique solution is $\chi(\sigma) = e^{\sigma L}\psi(x)$. Substituting the definition of $\chi$ yields \eqref{eq:exactjump}.
\end{proof}
Lemma~\ref{lem:jump} shows that exact lifting of spike-induced jumps requires only $\mathcal{L}_B$-invariance, whereas exact Koopman representations for continuously actuated systems typically require invariance under both $\mathcal{L}_f$ and $\mathcal{L}_B$. This distinction is central to the proposed framework: although obtaining finite-dimensional Koopman models for general control-affine systems is often challenging, the impulsive nature of neuromorphic actuation allows the jump dynamics to be represented exactly in lifted coordinates under the substantially weaker condition~\eqref{eq:LBinv}.

The exact jump representation of Lemma~\ref{lem:jump} combines with the lifted drift model of Section~\ref{sec:setting} to yield the following lifted description of~\eqref{eq:plant}. \begin{proposition}[Lifted model]\label{prop:lifted} Suppose that the dictionary span is $\mathcal{L}_B$-invariant with matrix $L$. Let $A\in\mathbb{R}^{N\times N}$ be any lifted drift matrix (e.g., obtained from~\eqref{eq:gedmd}) and let $r$ be the residual~\eqref{eq:res}. Then, the lifted state $y(t)=\psi(x(t))$ satisfies \begin{equation}\label{eq:lifted} \begin{aligned} \dot y(t)&=Ay(t)+r(x(t)), && t\neq t_i,\\ y(t_i^+)&=e^{s_i\alpha L}y(t_i^-), && i\ge1 . \end{aligned} \end{equation}
The jump dynamics are exact. Moreover, $r\equiv0$ if and only if the dictionary span is also invariant under $\mathcal{L}_f$, in which case \eqref{eq:lifted} is an exact linear impulsive representation of \eqref{eq:plant}. \end{proposition} \begin{proof} Between spikes, $\dot y=\mathcal{L}_f\psi(x)=Ay+r(x)$ by~\eqref{eq:res}. At each spike, Lemma~\ref{lem:jump} gives $y(t_i^+)=e^{s_i\alpha L}y(t_i^-)$. \end{proof}
Only the jump part of~\eqref{eq:lifted} is used in the sequel. The controller evaluates the lifted state $y(t_k)=\psi(x(t_k))$ from the measured state and predicts the effect of a spike with the exact matrices $e^{\pm\alpha L}$; the drift matrix $A$ and the residual $r$ appear neither in the control law nor in the stability analysis of Section~\ref{sec:stability}, which treats the flow on the original system. The drift enters only through the generator data used to learn the eigenfunctions and to synthesize the Lyapunov function in Section~\ref{sec:design}.

 Since \[ \mathcal{L}_B h(x) = \left.\frac{\mathrm d}{\mathrm d\sigma} h(x+\sigma B)\right|_{\sigma=0}, \] condition~\eqref{eq:LBinv} depends only on the variation of the observables along the direction $B$. Important examples include:
 \begin{itemize}
 \item[(i)] \emph{Polynomials:} all monomials of total degree at most $d$, since directional differentiation lowers degree.
 \item[(ii)] \emph{Products:} observables $g_\ell(x)(B^\top x)^k$, where $\mathcal{L}_B g_\ell\equiv0$ and each included power is accompanied by all lower powers. Then \[ \mathcal{L}_B\!\left(g_\ell(x)(B^\top x)^k\right) = k\|B\|^2 g_\ell(x)(B^\top x)^{k-1}. \]
 \item[(iii)] \emph{Exponential and trigonometric factors:} functions of type (ii) multiplied by $e^{\mu B^\top x}$ or by the pair $\cos(\omega B^\top x)$, $\sin(\omega B^\top x)$.
 \end{itemize}
 By the classical characterization of finite-dimensional translation-invariant function spaces~\citep{Engert1970}, case (iii) is essentially the most general finite-dimensional form: along $B$, the observables must be exponential polynomials. In cases (i) and (ii), $L$ is nilpotent, implying that $e^{\sigma L}$ is a finite matrix polynomial in $\sigma$. By contrast, generic radial basis functions do not possess a $\mathcal{L}_B$-invariant span.

We now construct a lifted Lyapunov function and the corresponding greedy firing rule. Assume $\psi_1(x)\equiv1$, define \[ \bar\psi(x):= (\psi_2(x),\ldots,\psi_N(x))^\top, \] and let $ \Pi:= \begin{bmatrix} -\bar\psi(0) & I_{N-1} \end{bmatrix} \in\mathbb{R}^{(N-1)\times N}$.  The centred lifted coordinates are  $z(x):=\Pi\psi(x) = \bar\psi(x)-\bar\psi(0)$, and we consider the quadratic Lyapunov function
\begin{equation}\label{eq:V}
V(y):= y^\top\Pi^\top P\Pi y, \qquad P\succ0.
\end{equation}
This induces the state-space function
\begin{equation}\label{eq:W}
W(x) := V(\psi(x)) = z(x)^\top Pz(x).
\end{equation}
Hence, $W(x(t))=V(y(t))$ along all trajectories. If $z(x)=0$ implies $x=0$, then $W$ is positive definite.

 \emph{Lifted greedy firing rule.} At tick $t_k$, the predicted change in $V$ produced by a spike $s\in\{-1,1\}$ is
\begin{equation}\label{eq:Delta}
\begin{aligned}
    \Delta_s(t_k) :=& V\!\left(e^{s\alpha L}y(t_k)\right)-V(y(t_k)) \\=& y(t_k)^\top \big(Q(s\alpha)-Q(0)\big) y(t_k)
\end{aligned}\,
\end{equation}
where \[ Q(\sigma) := e^{\sigma L^\top}\Pi^\top P\Pi\,e^{\sigma L}. \] Following the greedy strategy of \citet{Heemels2026EJC}, the controller fires the spike producing the largest decrease, provided that the decrease exceeds a prescribed cost $c\ge0$:
\begin{equation}\label{eq:greedy}
s(t_k)= \begin{cases} s^\star(t_k), & \Delta_{s^\star(t_k)}(t_k)\le -c, \\ 0, & \text{otherwise},
\end{cases}
\end{equation}
where  $s^\star(t_k) := \arg\min_{s\in\{-1,1\}} \Delta_s(t_k)$.  The parameter $c$ suppresses marginally beneficial spikes and determines the performance--spiking trade-off. The rule requires only the lifted state and the constant matrices $Q(\pm\alpha)-Q(0)$. For the linear dictionary $\psi(x)=(1,x^\top)^\top$, it reduces to the greedy controller of \citet{Heemels2026EJC}.

\begin{proposition}[Exactness of the lifted firing rule] \label{prop:test}
Suppose that the dictionary span is $\mathcal{L}_B$-invariant with matrix $L$. Then, for all $x\in\mathcal X$ and $\sigma\in\mathbb{R}$,
\begin{equation}\label{eq:test}
\begin{aligned} V\big(e^{\sigma L}\psi(x)\big) &= W(x+\sigma B)\\ &= \psi(x)^\top Q(\sigma)\psi(x). \end{aligned} \end{equation}
Consequently, \[ \Delta_s(t_k) = W(x(t_k)+s\alpha B) - W(x(t_k)), \] so the predicted decrease exactly equals the true decrease in $W$. Hence \eqref{eq:greedy} selects precisely the spike that a greedy controller operating directly on $W$ would select.
\end{proposition}

\begin{proof}
Lemma~\ref{lem:jump} gives  $e^{\sigma L}\psi(x) = \psi(x+\sigma B)$.  Substituting into \eqref{eq:W} yields  $V(e^{\sigma L}\psi(x)) = W(x+\sigma B)$. The quadratic representation follows directly from \eqref{eq:V} and the definition of $Q(\sigma)$.
\end{proof}
Proposition~\ref{prop:test} is dictionary independent and holds for any $\mathcal{L}_B$-invariant span. Its significance is that the greedy decision is exact despite operating entirely in lifted coordinates. If the jump representation were only approximate, the predicted and actual changes in $W$ could differ, and a fired spike might increase the Lyapunov function.

\begin{remark}[Neuromorphic implementation] The proposed rule preserves the structure of the linear greedy controller: precomputed matrices, quadratic evaluations, and linear updates. Following a spike, an encoded lifted state can be updated through the constant matrices $e^{\pm\alpha L}$ without re-evaluating the nonlinear observables. Only the current lifted state $y(t_k)=\psi(x(t_k))$ must be computed from the measured state at each tick. \end{remark}

\section{Stability Analysis} \label{sec:stability}

This section establishes practical stability of the lifted greedy controller~\eqref{eq:greedy} under the clock constraint $\tau$. The analysis considers one clock interval $[t_k,t_{k+1}]$: the effect of the spike at $t_k$ and the subsequent drift evolution are bounded separately and then combined through a control-Lyapunov condition. For $R>0$, define the Lyapunov sublevel set $\Omega_R:=\{x\in\mathcal X: W(x)\le R\}$.

\begin{assumption}[Certificate]\label{ass:cert}
The drift $f$ is continuously differentiable, and the function $W$ in~\eqref{eq:W} is twice continuously differentiable and positive definite. There exist $R>0$ and $\beta\in(0,1/\tau)$ such that $\Omega_R$ is compact, the flow $\Phi_\vartheta(x)$ exists for all $x\in\Omega_R$ and $\vartheta\in[0,\tau]$, and
\begin{equation}\label{eq:H}
\mathcal{L}_f W(x)-\bar u\,|\mathcal{L}_B W(x)| \le -\beta W(x), \qquad x\in\Omega_R .
\end{equation}
Furthermore, there exists a compact set $\Omega^+\supseteq\Omega_R$ containing all points $x+\sigma B$ and $\Phi_\vartheta(x)$ with $x\in\Omega_R$, $|\sigma|\le\alpha$, and $\vartheta\in[0,\tau]$.
\end{assumption}

Condition~\eqref{eq:H} is a bounded-input control-Lyapunov condition~\citep{LinSontag1991} for the continuously actuated system $\dot x=f(x)+Bv$ with $|v|\le\bar u=\alpha/\tau$, since \[ \min_{|v|\le\bar u} \bigl(\mathcal{L}_f W(x)+v\mathcal{L}_B W(x)\bigr) = \mathcal{L}_f W(x)-\bar u|\mathcal{L}_B W(x)|. \] Thus, the certificate characterizes the stabilizing authority available from the maximum average actuation that can be delivered by one spike per clock period. The set $\Omega_R$ is the certified operating region, whereas $\Omega^+$ bounds all states encountered during jump and flow transients over a single clock interval. A minimal choice is
\begin{equation}\label{eq:Omplus} \begin{aligned}
\Omega^+_{\min}:=\ & \{x+\sigma B:\ x\in\Omega_R,\ |\sigma|\le\alpha\} \\ &\cup \{\Phi_\vartheta(x):\ x\in\Omega_R,\ \vartheta\in[0,\tau]\},
\end{aligned}
\end{equation}
although any compact superset may be used in practice, e.g., a larger sublevel set $\{W\le\mu R\}$ with $\mu>1$.

The closed-loop evolution over one tick is as follows. At time $t_k$, the state lies in $\Omega_R$ and the greedy firing decision preserves this property. During $(t_k,t_{k+1})$, the drift may temporarily move the state outside $\Omega_R$, but it remains within $\Omega^+$. The subsequent analysis shows that the state returns to $\Omega_R$ before the next control decision and ultimately converges to an explicit practical bound. The role of $\Omega^+$ is therefore purely technical: it bounds the inter-sample excursion required to estimate the drift effects.

As the input is quantized in both amplitude and time, asymptotic stabilization cannot generally be achieved. Fixed-amplitude spikes, fired at most once per clock period, cannot compensate arbitrarily small state deviations. The appropriate notion is therefore \emph{practical stability}, whereby trajectories remain bounded and ultimately converge to a neighborhood of the origin rather than to the origin itself~\citep{LaSalle1961,Lakshmikantham1990}. This notion is closely related to uniform ultimate boundedness \citep[Sec.~4.8]{Khalil2002}. We adopt the following set-based definition.

\begin{definition}[Practical stability] \label{def:ps}
Let $\Omega_0$, $\Omega_1$, and $\Omega_{\mathrm u}$ be compact subsets of $\mathcal X$ satisfying  $0\in\Omega_{\mathrm u} \subseteq \Omega_1$, $ \Omega_0\subseteq\Omega_1$. The closed-loop system is \emph{practically stable with respect to} $(\Omega_0,\Omega_1,\Omega_{\mathrm u})$ if every solution with $x(0)\in\Omega_0$ satisfies:
\begin{itemize}
\item[(i)] $x(t)\in\Omega_1$ for all $t\ge0$;
\item[(ii)] there exists $T\ge0$ such that $x(t)\in\Omega_{\mathrm u}$ for all $t\ge T$.
\end{itemize}
The sets $\Omega_0$, $\Omega_1$, and $\Omega_{\mathrm u}$ are referred to as the admissible initial set, transient bound, and ultimate set, respectively. If a design parameter can be chosen so that $\Omega_{\mathrm u}$ is contained in an arbitrary neighborhood of the origin while $\Omega_0$ remains fixed, the closed loop is said to be \emph{practically stable with an arbitrarily small ultimate set}.
\end{definition}

Next, Theorem~\ref{thm:main} will establish explicit bounds on $W$ that imply practical stability (Corollary~\ref{cor:ps}) and show that the ultimate set can be made arbitrarily small by decreasing the clock period while keeping the average actuation level $\bar u=\alpha/\tau$ fixed (Corollary~\ref{cor:tau}). We first derive bounds on the effect of the firing decision and the subsequent drift evolution.

\begin{lemma}[Effect of the firing decision]\label{lem:jumpW}
Let $W$ be twice continuously differentiable and $\Omega_R$ compact. Define $D(t_k):=\mathcal{L}_B W(x(t_k))$  and \[ \kappa_2:= \sup_{x\in\Omega_R,\;|\sigma|\le\alpha} \left| \frac{\mathrm d^2}{\mathrm d\sigma^2} W(x+\sigma B) \right|. \] For any $x(t_k)\in\Omega_R$: \begin{itemize}
\item[(a)] $W(x(t_k^+))\le W(x(t_k))$, and hence $x(t_k^+)\in\Omega_R$;
\item[(b)] if a spike is fired,
\[ \begin{aligned} &W(x(t_k^+))-W(x(t_k))\\ &\quad\le \min\!\left\{ -c,\, -\alpha|D(t_k)|+\tfrac12\alpha^2\kappa_2 \right\}; \end{aligned} \]
\item[(c)] if no spike is fired, \[ \alpha|D(t_k)| < \tfrac12\alpha^2\kappa_2+c; \] \item[(d)] in all cases,
\begin{equation}\label{eq:step1}
W(x(t_k^+))-W(x(t_k)) \le -\alpha|D(t_k)| +\tfrac12\alpha^2\kappa_2+c . \end{equation}
\end{itemize}
\end{lemma}
\begin{proof} Property (a) follows directly from the firing rule, since either a spike decreases $W$ by at least $c$ or no spike is applied. Let \[ g(\sigma):=W(x(t_k)+\sigma B), \qquad |\sigma|\le\alpha . \] Then $g'(0)=D(t_k)$, $|g''(\sigma)|\le\kappa_2$.  By Proposition~\ref{prop:test}, \[ \Delta_s(t_k)=g(s\alpha)-g(0), \qquad s\in\{-1,1\}, \] so the greedy rule selects \[ s^\star(t_k)=\arg\min_{s=\pm1}g(s\alpha), \] fires if $g(s^\star\alpha)\le g(0)-c$, and otherwise sets $s(t_k)=0$. Let $\bar s:=-\mathrm{sgn}(D(t_k))$, with $\bar s=1$ when $D(t_k)=0$. Taylor's theorem gives \[ g(\bar s\alpha) = g(0) -\alpha|D(t_k)| + R, \] where \[ |R| \le \kappa_2 \int_0^\alpha (\alpha-r)\,\mathrm dr = \tfrac12\alpha^2\kappa_2 . \] Hence \begin{equation}\label{eq:sbar}
g(\bar s\alpha) \le g(0)-\alpha|D(t_k)|+\tfrac12\alpha^2\kappa_2.
\end{equation} Since $s^\star$ minimizes $g(s\alpha)$, \begin{equation}\label{eq:sstar}
g(s^\star\alpha)\le g(\bar s\alpha).
\end{equation}
If a spike is fired, \[ W(x(t_k^+))-W(x(t_k)) = g(s^\star\alpha)-g(0), \] which is bounded by both $-c$ (firing condition) and $-\alpha|D(t_k)|+\frac12\alpha^2\kappa_2$ from \eqref{eq:sbar}--\eqref{eq:sstar}, proving (b). If no spike is fired, then $g(s^\star\alpha)>g(0)-c$. Combining this with \eqref{eq:sbar}--\eqref{eq:sstar} yields (c). Finally, (d) follows from (b) when a spike is fired and from (c) when no spike is fired. \end{proof}

 While Lemma~\ref{lem:jumpW}(a) guarantees that the firing decision does not increase $W$, the stability proof requires the stronger estimate (d), which quantifies the reduction achieved along the input direction. The leading term $\alpha|D(t_k)|=\tau\bar u|\mathcal{L}_B W(x(t_k))|$ later combines with the drift term of Lemma~\ref{lem:flowW} to recover the control-Lyapunov condition~\eqref{eq:H}. The threshold $c$ only determines whether a spike is worthwhile; it does not provide the stabilizing effect itself. Parts (b) and (c) show that the controller either obtains a decrease comparable to the first-order prediction $\alpha|D(t_k)|$, up to a curvature error, or remains silent only when the potential benefit is already small. Inequality~\eqref{eq:step1} summarizes both cases in a single bound.

  Lemma~\ref{lem:jumpW} concerns the decision instant only and requires only the smoothness of $W$ and finiteness of $\kappa_2$. The next result bounds the drift evolution over one clock interval.

  \begin{lemma}[Flow over one tick]\label{lem:flowW} Let Assumption~\ref{ass:cert} hold, except possibly \eqref{eq:H}, and let $x(t_k)\in\Omega_R$. Then:
  \begin{itemize}
  \item[(a)] $x(t)\in\Omega^+$, and  $W(x(t)) \le W(x(t_k^+))+(t-t_k)m_0$, for all $t\in(t_k,t_{k+1}]$;
  \item[(b)]   and
  \begin{equation}\label{eq:step2}
  \begin{aligned} &W(x(t_{k+1}))-W(x(t_k^+)) \\
  &\qquad\le \tau\,\mathcal{L}_f W(x(t_k)) +\alpha\tau m_1 +\tfrac12\tau^2m_2 . \end{aligned}
  \end{equation}
  \end{itemize}
  where
  \begin{equation}\label{eq:consts}
  \begin{aligned}
  m_0&:= \sup_{x\in\Omega^+} |\mathcal{L}_f W(x)|, \\ m_1&:= \sup_{x\in\Omega^+} |\mathcal{L}_B\mathcal{L}_f W(x)|, \\ m_2&:= \sup_{x\in\Omega^+} |\mathcal{L}_f\mathcal{L}_f W(x)|.
  \end{aligned}
  \end{equation}
  \end{lemma}

  \begin{proof}
  By Lemma~\ref{lem:jumpW}(a), $x(t_k^+)\in\Omega_R$. Hence, Assumption~\ref{ass:cert} ensures that the flow arc $\Phi_\vartheta(x(t_k^+))$, $\vartheta\in[0,\tau]$, exists and remains in $\Omega^+$. Define  $h(\vartheta) := W(\Phi_\vartheta(x(t_k^+)))$.  Since $f\in C^1$ and $W\in C^2$, \[ \begin{aligned} h'(\vartheta) &= \mathcal{L}_f W(\Phi_\vartheta(x(t_k^+))),\\ h''(\vartheta) &= \mathcal{L}_f\mathcal{L}_f W(\Phi_\vartheta(x(t_k^+))). \end{aligned} \]
  Therefore $|h'(\vartheta)|\le m_0$, $ |h''(\vartheta)|\le m_2$. Part (a) follows immediately from \[ h(\vartheta) = h(0)+\int_0^\vartheta h'(r)\,\mathrm dr \] and the bound on $h'$. For part (b), integrating $h'(\vartheta)\le h'(0)+\vartheta m_2$ yields \[ W(x(t_{k+1}))-W(x(t_k^+)) \le \tau\,\mathcal{L}_f W(x(t_k^+)) +\tfrac12\tau^2 m_2. \] It remains to compare $\mathcal{L}_f W(x(t_k^+))$ and $\mathcal{L}_f W(x(t_k))$. If no spike is fired, the two states coincide. Otherwise, define \[ q(\sigma) := \mathcal{L}_f W(x(t_k)+\sigma B), \] for $\sigma\in[0,s(t_k)\alpha]$. Since the jump segment is contained in $\Omega^+$, \[ |q'(\sigma)| = |\mathcal{L}_B\mathcal{L}_f W(x(t_k)+\sigma B)| \le m_1. \] The mean-value theorem therefore gives \[ \mathcal{L}_f W(x(t_k^+)) \le \mathcal{L}_f W(x(t_k)) +\alpha m_1. \] Substituting into the previous inequality establishes \eqref{eq:step2}.
  \end{proof}

We now state the main stability result.

\begin{theorem}[Practical stability of clocked KGSC] \label{thm:main}
Let Assumption~\ref{ass:cert} hold and suppose $\delta\le \beta\tau R$, where
\begin{equation}\label{eq:delta}
\delta:= c+\tfrac12\alpha^2\kappa_2 +\alpha\tau m_1 +\tfrac12\tau^2m_2 . \end{equation}
Consider the closed-loop system \eqref{eq:plant}--\eqref{eq:jump} with the greedy firing rule \eqref{eq:greedy} at the sampling instants $t_k=k\tau$. Then, for every $x(0)\in\Omega_R$,

\begin{itemize}
\item[(a)] $x(t_k)\in\Omega_R$,  and \[ W(x(t_k)) \le (1-\beta\tau)^kW(x(0)) +\frac{\delta}{\beta\tau} \bigl(1-(1-\beta\tau)^k\bigr); \]
\item[(b)] $W(x(t)) \le W(x(t_k)) +\tau m_0, t\in(t_k,t_{k+1}]$;
\item[(c)]  $\limsup_{t\to\infty}W(x(t)) \le \frac{\delta}{\beta\tau} +\tau m_0$;
\item[(d)] the firing rate is bounded by $1/\tau$, and the long-term average input magnitude does not exceed $\bar u$.
\end{itemize}
\end{theorem}

\begin{proof}
Let $x(t_k)\in\Omega_R$. Combining \eqref{eq:step1} and \eqref{eq:step2} yields
\[ \begin{aligned} &W(x(t_{k+1}))-W(x(t_k))\\ &\quad\le \tau\mathcal{L}_f W(x(t_k)) -\alpha|D(t_k)| +\delta . \end{aligned} \] Using $\alpha=\bar u\tau$ and $D(t_k)=\mathcal{L}_B W(x(t_k))$ gives
\[ \begin{aligned} &W(x(t_{k+1}))-W(x(t_k))\\ &\quad\le \tau \Big( \mathcal{L}_f W(x(t_k)) -\bar u |\mathcal{L}_B W(x(t_k))| \Big) +\delta . \end{aligned} \]
Since $x(t_k)\in\Omega_R$, condition~\eqref{eq:H} implies
\begin{equation}\label{eq:rec}
W(x(t_{k+1})) \le (1-\beta\tau)W(x(t_k)) +\delta ,
\end{equation}
with $0<1-\beta\tau<1$. If $W(x(t_k))\le R$, then \[ W(x(t_{k+1})) \le (1-\beta\tau)R+\delta \le R, \] since $\delta\le\beta\tau R$. Hence $\Omega_R$ is invariant at the sampling instants. Because $x(0)\in\Omega_R$, relation~\eqref{eq:rec} holds for all $k$. Iterating \eqref{eq:rec} gives \[ W(x(t_k)) \le (1-\beta\tau)^kW(x(0)) +\delta \sum_{j=0}^{k-1}(1-\beta\tau)^j . \] Evaluating the geometric series establishes (a). By Lemmas~\ref{lem:jumpW}(a) and~\ref{lem:flowW}(a), \[ W(x(t)) \le W(x(t_k^+)) +\tau m_0 \le W(x(t_k)) +\tau m_0, \] for all $t\in(t_k,t_{k+1}]$, proving (b). Letting $k\to\infty$ in (a) yields \[ \limsup_{k\to\infty} W(x(t_k)) \le \frac{\delta}{\beta\tau}. \] Combining this with (b) establishes (c). Finally, (d) follows directly from the clocked implementation, which allows at most one spike per tick, together with the average-input bound derived in Section~\ref{sec:setting}.
\end{proof}

Theorem~\ref{thm:main} shows that the closed-loop system behaves like a contractive discrete-time system with disturbance level $\delta$. The quantity $\delta$ represents the price paid for clocking, fixed spike amplitudes, and the firing threshold. When $\delta\le\beta\tau R$, the Lyapunov decrease dominates these errors, yielding invariance of $\Omega_R$ and convergence to an explicit ultimate bound. The following corollary recasts these estimates in terms of practical stability.

\begin{corollary}[Practical stability]\label{cor:ps}
Suppose the assumptions of Theorem~\ref{thm:main} hold. For any $\epsilon>0$, define \[ r_\epsilon:= \frac{\delta}{\beta\tau}+\tau m_0+\epsilon, \qquad \Omega_{\mathrm u}^{\epsilon} := \{x\in\Omega^+:\ W(x)\le r_\epsilon\}. \]
Then, the closed-loop system is practically stable with respect to $(\Omega_R,\Omega^+,\Omega_{\mathrm u}^{\epsilon})$ in the sense of Definition~\ref{def:ps}. Moreover, \[ T= \tau \left\lceil \frac{\ln(\epsilon/R)} {\ln(1-\beta\tau)} \right\rceil \] is a valid settling-time estimate when $\epsilon<R$, whereas $T=0$ when $\epsilon\ge R$.
\end{corollary}

\begin{proof}
By Theorem~\ref{thm:main}(a) and Lemma~\ref{lem:flowW}(a), $x(t)\in\Omega^+,  \forall t\ge0$, establishing the transient bound. Let $k^\star$ be the smallest integer satisfying $(1-\beta\tau)^{k^\star}R\le\epsilon$. Then \[ k^\star= \left\lceil \frac{\ln(\epsilon/R)} {\ln(1-\beta\tau)} \right\rceil \] for $\epsilon<R$, and $k^\star=0$ otherwise. Since $W(x(0))\le R$, Theorem~\ref{thm:main}(a) gives \[ W(x(t_k)) \le \frac{\delta}{\beta\tau}+\epsilon, \qquad k\ge k^\star . \] Combining with Theorem~\ref{thm:main}(b), \[ W(x(t)) \le \frac{\delta}{\beta\tau} +\tau m_0+\epsilon = r_\epsilon, \qquad t\ge k^\star\tau , \] and therefore $x(t)\in\Omega_{\mathrm u}^{\epsilon}$ for all $t\ge T:=k^\star\tau$. \end{proof}

Since $W$ is continuous and positive definite, the ultimate set $\Omega_{\mathrm u}^{\epsilon}$ shrinks to the origin as $\delta/(\beta\tau)+\tau m_0+\epsilon \to 0$. The next result shows that this quantity can be made arbitrarily small.

\begin{corollary}\label{cor:tau}
Fix $\bar u$ and $\bar\tau>0$. Suppose Assumption~\ref{ass:cert} holds for $\bar\alpha=\bar u\bar\tau$, with constants $\kappa_2$, $m_0$, $m_1$, and $m_2$ computed on the corresponding set $\Omega^+$. For any $\tau\le\bar\tau$, let  $\alpha=\bar u\tau$, $c\le c_0\tau^2$, for some $c_0\ge0$. Then \[ \delta \le \tau^2(c_0+K_\delta), \] where $K_\delta := \tfrac12\bar u^2\kappa_2 +\bar u m_1 +\tfrac12 m_2$.
Consequently, the condition $\delta\le\beta\tau R$ of Theorem~\ref{thm:main} is satisfied whenever \begin{equation}\label{eq:taumax} \tau \le \min\{\bar\tau,\tau_{\max}\}, \qquad \tau_{\max} := \frac{\beta R}{c_0+K_\delta}. \end{equation} Moreover, \[ \frac{\delta}{\beta\tau}+\tau m_0 \le \tau \left( \frac{c_0+K_\delta}{\beta} +m_0 \right), \] which converges to zero as $\tau\to0$. Hence the closed-loop system is practically stable with an arbitrarily small ultimate set.
\end{corollary}

\begin{proof}
For $\tau\le\bar\tau$, it follows $\alpha=\bar u\tau\le\bar\alpha$. Hence, the corresponding jump segments and one-tick flow arcs are contained in those associated with $\bar\tau$, so the same set $\Omega^+$ remains valid and the constants $\kappa_2$, $m_0$, $m_1$, and $m_2$ do not increase. Since $\bar u$ is fixed, condition~\eqref{eq:H} is unchanged. Substituting $\alpha=\bar u\tau$ and $c\le c_0\tau^2$ into~\eqref{eq:delta} yields \[ \delta \le \tau^2 \left( c_0 +\tfrac12\bar u^2\kappa_2 +\bar u m_1 +\tfrac12m_2 \right) = \tau^2(c_0+K_\delta). \]
Therefore, $\delta\le\beta\tau R$ whenever $\tau\le\tau_{\max}$. The bound on the ultimate level follows immediately by dividing by $\beta\tau$.
\end{proof}

\section{Eigenfunction-based design}\label{sec:design}
Assumption~\ref{ass:cert} does not prescribe how the matrix $P$ is chosen. A linear design would derive $P$ from a Riccati equation of a local linearisation, which may not capture the global nonlinear structure of the system. Instead, we construct $W$ from Koopman eigenfunctions. Eigenfunctions associated with stable modes decay naturally under the drift and therefore require no control effort, whereas neutral and unstable modes must be actively regulated. Let \[ \varphi_j(x)=v_j^\top z(x), \qquad j\in\mathcal J, \] be (approximate) Koopman eigenfunctions corresponding to the neutral and unstable modes, with coefficient vectors $v_j\in\mathbb R^{N-1}$. We define
\begin{equation}\label{eq:Weig} W(x) = \sum_{j\in\mathcal J} w_j\,\varphi_j(x)^2 +\varepsilon\|z(x)\|^2,
\end{equation}
where $w_j>0$ and $\varepsilon>0$ ensures positive definiteness. This corresponds to~\eqref{eq:W} with \[ P = \sum_{j\in\mathcal J} w_j v_jv_j^\top +\varepsilon I_{N-1}. \]
Hence, the results of Proposition~\ref{prop:test} and Theorem~\ref{thm:main} apply directly. The weights are synthesised in Section~\ref{sec:synth}. The motivation for emphasizing unstable eigenfunctions is given by the following standard observation \citep[cf.][]{Mauroy2016}.

\begin{proposition}\label{prop:stable}
Let $\varphi_u$ be a continuous Koopman eigenfunction satisfying \[ \mathcal{L}_f\varphi_u(x)=\lambda_u\varphi_u(x), \qquad \lambda_u>0, \] on a forward-invariant neighbourhood $\mathcal V$ of the equilibrium. If an unforced trajectory $x(t)=\Phi_t(x(0))$ converges to the equilibrium, then $\varphi_u(x(0))=0$. Consequently, the local stable manifold is contained in the zero set of $\varphi_u$.
\end{proposition}

\begin{proof} Along unforced trajectories, \[ \varphi_u(x(t)) = e^{\lambda_u t}\varphi_u(x(0)). \] Since $x(t)\to0$, continuity implies $\varphi_u(x(t))\to\varphi_u(0)$. Moreover, $\lambda_u\varphi_u(0)=\mathcal{L}_f\varphi_u(0)=0$, so $\varphi_u(0)=0$. Because $e^{\lambda_u t}\to\infty$, necessarily $\varphi_u(x(0))=0$. \end{proof}

Penalizing unstable eigenfunctions therefore drives the state toward the stable manifold, where the drift dynamics guide it to the equilibrium without further actuation. Neutral eigenfunctions capture the dominant global structure of the dynamics. For conservative mechanical systems, the total energy is a neutral eigenfunction; with weak damping it is approximately conserved, and the regression below returns the most nearly conserved observable in the span. Penalizing its deviation from the equilibrium energy recovers the principle of energy-based control~\citep{AstromFuruta2000}.

\subsection{Learning the eigenfunctions from data}

Let $x^{(j)}$, $j=1,\ldots,M$, be samples from unforced trajectories, and let $x^{(j,-)}$ and $x^{(j,+)}$ denote samples taken a time $h$ before and after $x^{(j)}$ on the same trajectory. Form the matrices $Z\in\mathbb{R}^{M\times(N-1)}$ with rows $z(x^{(j)})^\top$ and $\dot Z\in\mathbb{R}^{M\times(N-1)}$ with rows \[ \frac{\big(\bar\psi(x^{(j,+)})-\bar\psi(x^{(j,-)})\big)^\top}{2h}, \] which provide finite-difference estimates of the generator data~\citep{Klus2020}.
\begin{itemize}
\item \emph{Neutral eigenfunction ($\lambda=0$).} Compute \[ v_0 = \arg\min_{\|v\|=1} \|\Gamma_0\dot Z\,v\|, \] where \[ \Gamma_0 = \mathrm{diag} \!\left( \frac{1}{1+\|z(x^{(j)})\|^2} \right). \] Equivalently, $v_0$ is the right singular vector of $\Gamma_0\dot Z$ associated with its smallest singular value. The resulting vector is subsequently rescaled.

\item \emph{Unstable eigenfunction.} The unstable eigenvalue $\lambda_u$ is estimated first, as the positive eigenvalue of a linear model $\dot x\approx\hat A x$ fitted by least squares to the samples near the equilibrium. Then \[ v_u = \arg\min_{v:\,v_{\iota}=1} \| \Gamma_u(\dot Z-\lambda_u Z)v \|, \] where the coefficient $v_\iota$ of the coordinate that dominates the linear unstable eigenfunction is fixed to one, which excludes $v=0$ and fixes the scale. Here $\Gamma_u = \mathrm{diag}\big(\gamma(x^{(j)})^{1/2}\big)$ with Gaussian locality weights $\gamma(x)=\exp(-\|x\|_S^2/\ell^2)$ centered at the equilibrium, for a scaling matrix $S$ and a width $\ell$. The locality is needed because the eigen-equation can be satisfied accurately in the span only near the equilibrium. This amounts to the weighted eigen-regression of \citet{Kaiser2021}.
\end{itemize}
Both computations reduce to linear least-squares problems and require only the eigenfunctions of interest rather than a full Koopman-model identification, thereby avoiding spurious eigenvalues caused by model truncation.

\subsection{Convex synthesis of the Lyapunov matrix}\label{sec:synth}

The Koopman eigenfunctions determine the \emph{structure} of $W$, but their weights remain to be chosen. A linear design would obtain $P$ from a Riccati inequality of the linearized dynamics, yielding a quadratic control Lyapunov function. Here, we instead synthesize $P$ directly from the nonlinear certificate~\eqref{eq:H}. For $ W(x)=z(x)^\top Pz(x)$, the quantities appearing in~\eqref{eq:H} are
\begin{equation}\label{eq:LfWLBW}
\begin{aligned}
\mathcal{L}_f W(x)&=2z(x)^\top P\,\mathcal{L}_f z(x),\\
\mathcal{L}_B W(x)&=2z(x)^\top P\,\Pi L\psi(x),
\end{aligned}
\end{equation}
where $\mathcal{L}_f z(x)=\nabla\bar\psi(x)f(x)$ is obtained either from the model or from finite-difference estimates, while $\Pi L\psi(x)$ is known exactly. For a linear system with $\psi(x)=(1,x^\top)^\top$, condition~\eqref{eq:H} reduces to \[ x^\top(A^\top P+PA)x - 2\bar u|B^\top Px| \le -\beta x^\top Px, \] which is the rate-limited counterpart of the greedy-control condition in \citet{Heemels2026EJC}.

Since~\eqref{eq:H} is nonconvex because of the absolute value, we introduce a nominal feedback $\kappa:\mathcal X\to[-\bar u,\bar u]$ and require \begin{equation}\label{eq:S}
\begin{aligned}
2z(x)^\top P\big(\mathcal{L}_f z(x)+\kappa(x)\Pi L\psi(x)\big) &\\+\beta z(x)^\top Pz(x) \le -\eta\|z(x)\|^2 ,
\end{aligned}
\end{equation}
for some margin $\eta>0$.

\begin{proposition}[Convex sufficient condition] \label{prop:S}
For fixed $\kappa$ satisfying $|\kappa(x)|\le\bar u$, condition~\eqref{eq:S} is linear in $P$. Moreover, if~\eqref{eq:S} holds on a set $\mathcal D$, then \eqref{eq:H} also holds on $\mathcal D$.
\end{proposition}

\begin{proof} By~\eqref{eq:LfWLBW}, \[ \mathcal{L}_f W(x)+\kappa(x)\mathcal{L}_B W(x)+\beta W(x) \le -\eta\|z(x)\|^2 . \] Since \[ \kappa(x)\mathcal{L}_B W(x) \ge -\bar u|\mathcal{L}_B W(x)|, \] condition~\eqref{eq:H} follows immediately.
\end{proof}

The function $\kappa$ is not applied to the plant and the controller always uses the greedy spiking law~\eqref{eq:greedy}; $\kappa$ serves only as an auxiliary witness that removes the absolute value from \eqref{eq:H}. By Proposition~\ref{prop:S}, any admissible $\kappa$ satisfying $|\kappa(x)|\le\bar u$ yields a valid certificate. The relaxation is exact at $x$ when \[ \kappa(x)\mathcal{L}_B W(x) = -\bar u |\mathcal{L}_B W(x)|, \] i.e., when $\kappa$ equals the bounded input that maximally decreases $W$,
\begin{equation}\label{eq:kstar}
\begin{aligned}
    \kappa_W^\star(x) =& -\bar u\,\mathrm{sgn}\!\big(\mathcal{L}_B W(x)\big) \\&\in \arg\min_{|v|\le\bar u} \bigl(\mathcal{L}_f W(x)+v\mathcal{L}_B W(x)\bigr).
\end{aligned}
\end{equation}
Since $\kappa_W^\star$ depends on the unknown matrix $P$, we compute $\kappa$ and $P$ iteratively. Toward this, we parameterize \begin{equation}\label{eq:Pparam} \begin{aligned}
P=&\, w_0v_0v_0^\top +w_uv_uv_u^\top +\operatorname{diag}(d) +\varepsilon I_{N-1},\\ &w_0,w_u\ge0,\qquad d\in\mathbb{R}_{\ge0}^{N-1},
\end{aligned}
\end{equation}
which guarantees $P\succ0$ and makes all constraints linear in the decision variables. Let \[ \mathcal T_\rho = \{x:x^\top P_{\rm ref}x\le\rho\}, \qquad \mathcal D = \mathcal T_{\gamma\rho}, \quad \gamma>1. \] Given interior samples $x^{(i)}\in\mathcal D$, boundary samples $x^{(b)}\in\partial\mathcal D$, and target samples $x^{(t)}\in\mathcal T_\rho$, we solve \begin{equation}\label{eq:LP}
\begin{aligned}
\min_{w_0,w_u,d}\quad &\sum_i W(x^{(i)})\\
\text{s.t.}\quad &\eqref{eq:S}\ \text{at all }x^{(i)},\\
&W(x^{(b)})\ge1, \qquad W(x^{(t)})\le1. \end{aligned}
\end{equation}
The boundary constraint ensures that the certified sublevel set $\{W\le1\}$ remains inside $\mathcal D$, whereas the target constraint enforces $\mathcal T_\rho\subseteq\{W\le1\}$. The largest feasible $\rho$ is found by bisection. As in sampling-based synthesis of control Lyapunov functions~\citep{Ravanbakhsh2019}, the constraints hold only at the samples, so the resulting certificate is verified a posteriori on a dense grid; for polynomial dictionaries and drifts, sum-of-squares constraints can replace the samples~\citep{Parrilo2003}.

\emph{Preserving the neutral modes.} Program~\eqref{eq:LP} involves only samples of a neighbourhood $\mathcal D$ of the equilibrium, whereas the neutral eigenfunctions are included in $W$ for their global role (Section~\ref{sec:design}). The next result shows that they are nearly invisible to a local certificate.

\begin{proposition} \label{prop:flat}
Let $f$ be continuously differentiable with $f(0)=0$ and $Df(0)$ nonsingular, and let $\varphi$ be twice continuously differentiable on a neighbourhood of the origin, with $\mathcal{L}_f\varphi\equiv0$ there and $\varphi(0)=0$. Then $\nabla\varphi(0)=0$; hence $\varphi(x)=O(\|x\|^2)$ and $\varphi(x)^2=O(\|x\|^4)$ as $x\to0$.
\end{proposition}

\begin{proof}
Differentiating $\nabla\varphi(x)^\top f(x)\equiv0$ gives $\nabla^2\varphi(x)f(x)+Df(x)^\top\nabla\varphi(x)=0$. At $x=0$, $f(0)=0$, so $Df(0)^\top\nabla\varphi(0)=0$, and $\nabla\varphi(0)=0$ because $Df(0)$ is nonsingular. Taylor's theorem with $\varphi(0)=0$ gives the order estimates. \end{proof}

By Proposition~\ref{prop:flat}, the terms $w_0\varphi_0^2$ are of fourth order near the equilibrium, whereas $\varphi_u^2$ and the regularisation are of second order; approximately neutral eigenfunctions inherit this property approximately. Program~\eqref{eq:LP} therefore has little incentive to retain $w_0$ and may lower it to enlarge the certified set, at the expense of the global behaviour that the samples in $\mathcal D$ do not represent. To keep this structural prior, we optionally impose the linear \emph{neutral-mode share} constraint
\begin{equation}\label{eq:ratio} w_0\ge r\,w_u, \qquad r\ge0, \end{equation}
with one such constraint per neutral mode if there are several. The choice $r=0$ recovers the unconstrained program. The parameter $r$ trades the size of the certified set against the preservation of the global structure encoded by the neutral modes, and is selected a posteriori from the closed-loop behaviour outside the certified set.

 To approximate the ideal witness~\eqref{eq:kstar}, we use the continuous feedback \begin{equation}\label{eq:kappaW} \kappa_W(x) = -\bar u\, \mathrm{sat}\!\left( \frac{\mathcal{L}_B W(x)}{\epsilon_s} \right), \end{equation} where $\mathrm{sat}(s)=\max\{-1,\min\{1,s\}\}$ and $\epsilon_s>0$ is a smoothing parameter. This satisfies $|\kappa_W|\le\bar u$ and coincides with $\kappa_W^\star$ whenever $|\mathcal{L}_B W(x)|\ge\epsilon_s$.

 \begin{proposition}\label{prop:kappa}
 For every $W$ and every $x$, \begin{equation}\label{eq:kloss}
 -\bar u|\mathcal{L}_B W(x)| \le \kappa_W(x)\mathcal{L}_B W(x) \le -\bar u|\mathcal{L}_B W(x)| +\tfrac14\bar u\epsilon_s .
 \end{equation}
 Hence, if the witness $\kappa=\kappa_W$ is built from the same $W$ that it certifies, the left-hand side of~\eqref{eq:S} exceeds that of~\eqref{eq:H} by at most $\frac14\bar u\epsilon_s$, and the two coincide whenever $|\mathcal{L}_B W(x)|\ge\epsilon_s$. \end{proposition}

 \begin{proof}
 The result follows directly from the definition of $\kappa_W$. For $|\mathcal{L}_B W|\ge\epsilon_s$, $\kappa_W=-\bar u\,\mathrm{sgn}(\mathcal{L}_B W)$ and equality holds. For $|\mathcal{L}_B W|<\epsilon_s$, $\kappa_W\mathcal{L}_B W=-\bar u(\mathcal{L}_B W)^2/\epsilon_s$, and the deviation $\bar u(|\mathcal{L}_B W|-(\mathcal{L}_B W)^2/\epsilon_s)$ is maximal at $|\mathcal{L}_B W|=\epsilon_s/2$, yielding $\frac14\bar u\epsilon_s$. \end{proof}
The constant bound matters away from the equilibrium. Near the equilibrium, where $|\mathcal{L}_B W|<\epsilon_s$, $\kappa_W=-(\bar u/\epsilon_s)\mathcal{L}_B W$ is a linear high-gain feedback, and~\eqref{eq:S} reduces there to a quadratic decrease condition that can be met with the margin $\eta\|z\|^2$.

As $\kappa_W$ depends on $P$, joint optimization of $(\kappa,P)$ would destroy convexity and we therefore alternate between the two:
\begin{equation}\label{eq:kappa}
\begin{aligned} \kappa_{m+1}&:=\kappa_{W_m},\\ W_{m+1}&:= \text{solution of~\eqref{eq:LP} with } \kappa=\kappa_{m+1},\end{aligned}
\end{equation}
starting from the eigenfunction design~\eqref{eq:Weig}. Each iterate yields a valid certificate by Proposition~\ref{prop:S}, whatever the quality of $\kappa_{m+1}$. Since $\kappa_{m+1}$ is built from $W_m$ but certifies $W_{m+1}$, conservatism remains where the signs of $\mathcal{L}_B W_m$ and $\mathcal{L}_B W_{m+1}$ differ; if the iteration converges to a fixed point, only the smoothing loss of Proposition~\ref{prop:kappa} remains. Monotone improvement of $\rho$ is not guaranteed, because the feasible set changes with $\kappa_{m+1}$; the bisection is therefore warm-started at the previous $\rho$, and the best certificate found is returned. The procedure is analogous to policy iteration or the $D$--$K$ iteration of robust control.  Algorithm \ref{alg:synth} summarizes the steps for the lifted Lyapunov matrix synthesis.

\begin{algorithm}[t] \caption{Convex synthesis of the lifted Lyapunov matrix} \label{alg:synth}
\begin{algorithmic}[1]
\small
\Require $v_0,v_u$, $P_{\mathrm{ref}}$, $\beta,\eta,\varepsilon,\gamma,\epsilon_s,\bar u$, $r$, $[\rho_{\mathrm{lo}},\rho_{\mathrm{hi}}]$, $m_{\max}$
\State Initialise $P_0$ from the eigenfunction design.
\For{$m=0,\ldots,m_{\max}-1$}
\State Compute $\kappa_{m+1}=\kappa_{W_m}$.
\State Perform bisection on $\rho$ and solve~\eqref{eq:LP}, with~\eqref{eq:ratio} if $r>0$.
\State Set $P_{m+1}$ to the solution at the largest feasible $\rho$; keep the best $(P,\rho)$ found so far; stop if~\eqref{eq:LP} is infeasible.
\EndFor
\State Verify~\eqref{eq:H} on a dense grid and compute $\Omega_R$ and $\tau_{\max}$.
\end{algorithmic} \end{algorithm}

\subsection{KGSC algorithm}

The complete procedure is summarized as follows. \noindent\emph{Offline:} (1) choose an $\mathcal{L}_B$-invariant dictionary and compute $L$; (2) identify the neutral and unstable eigenfunctions; (3) synthesize $P$ via Algorithm \ref{alg:synth};  (4) precompute $Q(\pm\alpha)-Q(0)$. \noindent\emph{Online:} at each tick $t_k$, evaluate $y(t_k)=\psi(x(t_k))$, compute $\Delta_{\pm1}(t_k)$ from~\eqref{eq:Delta}, and select $s(t_k)$ according to~\eqref{eq:greedy}.

\section{Case study: pendulum swing-up and stabilization with spikes}\label{sec:sim}
\subsection{Setup} We consider the inverted pendulum \begin{equation}\label{eq:pend}
\dot x_1=x_2,\qquad \dot x_2=a\sin x_1-bx_2+u,
\end{equation}
where $x_1\in\mathbb S^1$ is the angle measured from the upright equilibrium, $\mathbb S^1$ is the unit
circle (angles modulo $2\pi$), $x_2=\dot x_1$ is the angular velocity, $a=g/\ell=9.81\,\mathrm{s}^{-2}$, and $b=0.1\,\mathrm{s}^{-1}$. The linearization at the upright position has eigenvalues $3.08$ and $-3.18$, and the input direction is $B=(0,1)^\top$. Spikes induce velocity jumps $\Delta x_2=\pm\alpha=\pm0.06\,\mathrm{rad/s}$ and are generated at $100$ Hz ($\tau=10$ ms), giving the maximum average actuation $\bar u=\alpha/\tau=6\,\mathrm{rad/s^2}=0.61a$. Consequently, static balancing is impossible beyond $|x_1|=\arcsin(0.61)=0.66$ rad.

 Since spikes act only on $x_2$, we employ the $\mathcal{L}_B$-invariant family of Section~\ref{sec:lifting} with $g_\ell\in\{1,\sin x_1,\cos x_1\}$ and powers $k=0,1,2$ of $x_2$:
\begin{equation}\label{eq:dict} \begin{aligned}
\psi(x)=(&1,x_2,x_2^2,\sin x_1,x_2\sin x_1,x_2^2\sin x_1,\\ &\cos x_1,x_2\cos x_1,x_2^2\cos x_1)^\top .
\end{aligned}
\end{equation}
This yields $N=9$ observables. Since $\mathcal{L}_B(x_2^kg(x_1))=k\,x_2^{k-1}g(x_1)$,
\begin{equation}\label{eq:Lpend} L=I_3\otimes
\begin{bmatrix} 0&0&0\\ 1&0&0\\ 0&2&0
\end{bmatrix}, \qquad e^{\sigma L} = I_3\otimes
\begin{bmatrix} 1&0&0\\ \sigma&1&0\\ \sigma^2&2\sigma&1
\end{bmatrix}. \end{equation}
The span is not $\mathcal{L}_f$-invariant, so the flow model is only approximate, whereas the jump model remains exact.

 The design is learned from 300 unforced trajectories of length $2$ s, initialized uniformly in $[-\pi,\pi]\times[-8,8]$. This produces $3\times10^4$ samples with finite-difference step $h=1$ ms. The unstable eigenvalue is estimated by fitting $\dot x_2\approx\hat a\sin x_1-\hat b x_2$ near the equilibrium, the coefficient of $\sin x_1$ in $v_u$ is fixed to one, and $\gamma(x)=\exp(-\mathrm{wrap}(x_1)^2-(x_2/3)^2)$. Two KGSC designs are considered. The \emph{nominal} design uses $w_0=1$, $w_u=10$, and $\varepsilon=10^{-3}$ in \eqref{eq:Weig}. The \emph{synthesized} design solves \eqref{eq:LP} with $\beta=0.5$, $\eta=\varepsilon=10^{-3}$, $\epsilon_s=10^{-2}$, $\gamma=3$, approximately $4000$ interior, $500$ boundary, and $900$ target samples, ten bisection steps on $\rho$, three updates of \eqref{eq:kappa}, and $r\in\{0,0.03,0.05\}$ in~\eqref{eq:ratio}. Section~\ref{sec:cert} compares all designs. Since the near-global swing-up is the main result of the case study, Figs.~\ref{fig:basins} and~\ref{fig:swing} show the nominal design, which has the largest basin; the synthesized designs trade a slightly smaller basin for larger certificates (Table~\ref{tab:designs}). As a baseline, we use the linear greedy controller of \citet{Heemels2026EJC} with
 \[ W_{\rm lin}(x)=\tilde x^\top P_{\rm lin}\tilde x, \qquad \tilde x=(\mathrm{wrap}(x_1),x_2), \]
 where $P_{\rm lin}$ is obtained from the shifted Riccati equation of the linearized system. The spiking costs are chosen so that both controllers achieve comparable steady-state performance (about 5 and 4 spikes/s, with $|x_1|\le0.003$ and $0.004$\,rad, respectively). Simulations use a fourth-order Runge--Kutta solver with step size $1$ ms. A run is considered successful if $|\mathrm{wrap}(x_1)|<0.1$ and $|x_2|<0.5$ at $t=20$ s.

\subsection{Learned eigenfunctions and exact lifting} The learned neutral and unstable eigenfunctions are
\begin{align*} \varphi_0(x) =& \tfrac12x_2^2 +9.8119(\cos x_1-1)\\& -0.0037\sin x_1 +0.0042x_2\sin x_1,\\ \varphi_u(x) =& \sin x_1 +0.3129x_2 +0.0129x_2^2\sin x_1 \\&+0.0057x_2\cos x_1+\cdots,
\end{align*}
with estimated unstable eigenvalue $\lambda_u=3.0825$, equal to the true value to four digits. The function $\varphi_0$ recovers the mechanical energy $\frac12x_2^2+a(\cos x_1-1)$ with a $0.02\%$ error in $a$; since $b>0$ it is approximately rather than exactly conserved, $\frac{\mathrm d}{\mathrm dt}\varphi_0(x(t))\approx-b\,x_2(t)^2$. The function $\varphi_u$ refines the linear unstable eigenfunction $x_1+0.314x_2$ with periodic and nonlinear corrections, so that it is well defined on $\mathbb S^1$. Over 2000 random states and four values of $\sigma$, the lifted jump~\eqref{eq:exactjump} matches direct evaluation up to $2.8\times10^{-14}$, and~\eqref{eq:test} up to a relative error of $1.3\times10^{-14}$, i.e.\ to machine precision.

\subsection{Basins of attraction and swing-up} 
\begin{figure}[htbp]
\centering
\includegraphics[width=0.8\textwidth]{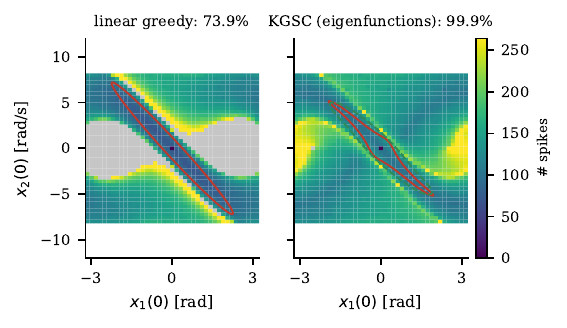}
\caption{Basins of attraction under the rate limit $\bar u=0.61a$ ($41\times41$ initial states, 20\,s).
Horizontal axis: initial angle from the upright position; vertical axis: initial angular velocity.
Colour: number of spikes until $t=20$\,s; grey: failure; red: certified set $\Omega_R$ of
Theorem~\ref{thm:main}; dashed green (right): certified set of the synthesized design of
Section~\ref{sec:synth} with $r=0$ (Table~\ref{tab:designs}).}
\label{fig:basins}
\end{figure}

\begin{figure}[htbp]
\centering
\includegraphics[width=0.62\textwidth]{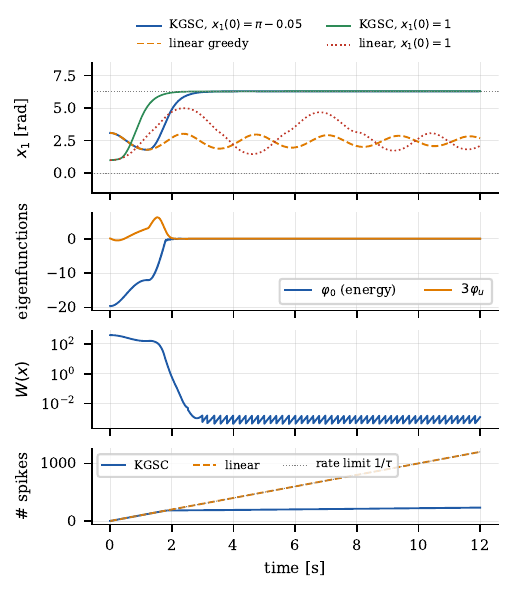}
\caption{Time responses. Top: angle $x_1(t)$, unwrapped, so that $0$ and $\pm2\pi$ are the upright
position, for KGSC and the linear greedy design from $x_1(0)=\pi-0.05$ and from $x_1(0)=1$. Then, for
KGSC from $\pi-0.05$: the eigenfunctions $\varphi_0(x(t))$ (energy) and $\varphi_u(x(t))$, the Lyapunov
function $W(x(t))$, and the cumulative spike counts, with the rate limit $t/\tau$ for reference.}
\label{fig:swing}
\end{figure}

Fig.~\ref{fig:basins} summarizes the results over a $41\times41$ grid of initial conditions. KGSC achieves a success rate of $99.94\%$ ($1680/1681$), including the hanging equilibrium. The only failure is a rate-saturation equilibrium near $x_1=-2.49$\,rad, where $a|\sin x_1|\approx\bar u$. In contrast, the linear greedy controller succeeds from only $73.9\%$ of the initial states and typically fails when gravity exceeds the available average actuation. Over the $1243$ initial conditions from which both controllers succeed, KGSC is also more spike-efficient, requiring on average $137$ spikes versus $148$ for the linear design. An ablation study further highlights the role of the neutral eigenfunction: setting $w_0=0$ reduces the success rate to $42.5\%$. Representative trajectories are shown in Fig.~\ref{fig:swing}. From the nearly hanging state $x_1(0)=\pi-0.05$, KGSC first regulates the energy and then suppresses the unstable mode, reaching the upright equilibrium within $2.8$\,s using $186$ spikes ($232$ spikes over the 12\,s response, including the steady state). The linear controller instead saturates the firing rate and remains near the downward equilibrium. Similarly, from $x_1(0)=1$\,rad, beyond the static holding limit, KGSC exploits the natural dynamics to swing the pendulum up and stabilize it, whereas the linear controller fails.

\subsection{Certificates and synthesized designs}\label{sec:cert}
Table~\ref{tab:designs} compares the designs. The certified set $\Omega_R$ is the largest sublevel set of $W$ on which~\eqref{eq:H} holds on a $801\times801$ grid, and $\tau_{\max}$ is the clock period certified by Corollary~\ref{cor:tau}, with the constants computed for $\bar\tau=10$\,ms and $c_0=0$. Both are conservative: the controller works at $\tau=10$\,ms, whereas the certified periods are of the order of 1\,ms.
\begin{itemize}
\item \emph{Without the share constraint} ($r=0$), the synthesis returns $w_0=0.12$, $w_u=8.35$ and a single diagonal weight $0.49$ on $(\cos x_1-1)^2$. The certified area grows by $80\%$ with respect to the nominal design and $\tau_{\max}$ by a factor of $2.9$, but the basin decreases to $97.4\%$. As anticipated by Proposition~\ref{prop:flat}, the energy $\varphi_0\approx\frac12(x_2^2-ax_1^2)$ is flat at the upright equilibrium, so the local program lowers its weight relative to $w_u$ from $0.1$ to $0.014$, which weakens the swing-up: the energy term matters precisely in the region far from the equilibrium, where the pendulum must be swung up. The bisection gives $\rho=2.963$, $2.970$ and $2.970$ in the three iterations, so almost all of the gain is obtained with the first witness $\kappa_{W_0}$.
\item \emph{With the share constraint}~\eqref{eq:ratio}, which for the pendulum preserves the share of the energy term, the trade-off is controlled by $r$. For $r=0.05$ the basin is $99.9\%$, equal to that of the nominal design up to one grid point, while the certified area is $30\%$ larger and $\tau_{\max}$ about $1.7$ times larger. For $r=0.03$, the certified area is $52\%$ larger and $\tau_{\max}$ about twice as large, with a basin of $99.6\%$.
\end{itemize}
The linear design has a large certified set around the upright position but the smallest basin, because the certificate is local and the linear Lyapunov function ignores the global geometry.

\begin{table}[htbp]
\centering
\caption{Comparison of the designs: certified area of $\Omega_R$ in the $(x_1,x_2)$ plane, certified clock period $\tau_{\max}$ (Corollary~\ref{cor:tau}), basin of attraction on the $41\times41$ grid and mean number of spikes of the successful runs ($\tau=10$\,ms, 20\,s). Linear greedy: the design of \citet{Heemels2026EJC}.}
\label{tab:designs}
\setlength{\tabcolsep}{3pt}

\begin{tabular}{@{}lrrrr@{}}
\toprule
Design & area & $\tau_{\max}$ [ms] & basin [\%] & spikes\\
\midrule
Linear greedy & 8.45 & 0.84 & 73.9 & 148\\
KGSC nominal & 5.13 & 0.67 & 99.94 & 151\\
Synthesized, $r=0$ & 9.21 & 1.97 & 97.4 & 167\\
Synthesized, $r=0.03$ & 7.79 & 1.39 & 99.6 & 161\\
Synthesized, $r=0.05$ & 6.69 & 1.10 & 99.9 & 175\\
\bottomrule
\end{tabular}
\end{table}

\section{Conclusions and Future Work}\label{sec:conclusion} This paper established a connection between Koopman lifting and neuromorphic greedy spiking control. For dictionaries whose span is invariant under differentiation along the input direction, spike-induced state translations admit an exact linear representation in lifted coordinates. Consequently, the firing decisions of greedy spiking controllers reduce to quadratic tests on the lifted state, while practical stability under firing-rate limitations can be certified through a bounded-input control-Lyapunov condition with explicit performance bounds. Building the Lyapunov function from learned neutral and unstable Koopman eigenfunctions yields a data-driven controller that combines energy regulation and local stabilization within a single spiking law. For the inverted pendulum, the resulting Koopman greedy spiking controller achieves near-global swing-up and stabilization under severe rate constraints, substantially outperforming a linear greedy design. Moreover, a linear-program synthesis procedure was developed to compute the Lyapunov weights through a convex relaxation of the rate-limited control-Lyapunov condition. It enlarges the certified region by up to $80\%$ and the certified clock period by up to a factor of three with respect to the hand-tuned weights, and, with a linear constraint that preserves the share of the neutral modes, it does so while keeping the swing-up basin of the hand-tuned design.

The results also highlight several directions for future work. The stability certificate remains local and does not formally certify the swing-up phase, while the current one-step greedy strategy can create rate-saturation equilibria. Extending the analysis to multi-step or predictive firing rules, reducing the conservatism of the stability bounds, and improving scalability through sparse or learned dictionaries are important next steps. More broadly, the framework naturally extends to multi-input and nonlinear control-affine systems through invariance with respect to the corresponding input operators, opening the way to Koopman-based predictive spiking control, and fully neuromorphic implementation.

\section*{Acknowledgment}
The author used the AI assistant Claude (Anthropic) as a research tool during this work: to check derivations, to implement and run the simulation code, and to draft text. The research problem, the overall approach and all editorial and technical decisions are the author's own. The author verified all results, including the proofs and the simulations, and takes full responsibility for the content of this article.

\end{document}